# Interfacial Energy of Copper Clusters in Fe-Si-B-Nb-Cu alloys

Rajesh Jha,[a] David R. Diercks,[b] Nirupam Chakraborti,[c] Aaron P. Stebner,[a,*] and Cristian V. Ciobanu[a,*]

[a]Department of Mechanical Engineering, Colorado School of Mines, Golden, Colorado 80401, USA
[b]Department of Metallurgical and Materials Engineering, Colorado School of Mines, Golden, Colorado 80401, USA
[c]Department of Metallurgical and Materials Engineering, Indian Institute of Technology Kharagpur, West Bengal, India

**Abstract:** Using a combination of numerical simulations and atom-probe tomography experiments, we determine the interfacial energy of Cu nanocrystals precipitated within the amorphous matrix of FINEMET (molar composition $Fe_{72.89}Si_{16.21}B_{6.90}Nb_3Cu_1$). Specifically, we use the Langer-Schwartz model implemented in the software Thermocalc to carry out parametric simulations of growth and coarsening of Cu clusters for different interface energies. We have carried out atom-probe tomography (APT) experiments to determine the interface energy as the value for which the simulated particle size distribution best matches the experimental data. This combination of APT and precipitation modeling can be applied to other nanocrystals precipitated within amorphous matrices.

Keywords: Interfacial energy, FINEMET, Atom Probe Tomography, Precipitation Kinetics, Thermocalc

*Corresponding Authors: A. P. Stebner (astebner@mines.edu) and C.V. Ciobanu (cciobanu@mines.edu )

The interfacial energy between precipitates and a matrix is an important property that determines many physical quantities of interest in precipitate-strengthened alloys − from the critical radius, activation energy, and nucleation rate in the early stages of the process, to the coarsening rates of the nanoparticles formed in the matrix. Hence, models of nucleation and growth use interface energy as a key parameter [1-8] to develop theoretical formulations for nucleation, growth, and coarsening. The early work of Wagner [1], Lifshitz and Slyozov [2], assumed sharp interfaces and derived a coarsening law by which the mean radius $r$ of a new phase nucleated in a matrix increases as the cube root of time $t$, $r(t) = r(0) + kt^{1/3}$, with the coarsening rate constant $k$ depending on interface energy $\gamma$ [9]; models based on diffuse interfaces [10, 11] lead to similar conclusions regarding the coarsening behavior. In the context of Ni-based super-alloys with γ' phase precipitates, experimental data from scanning electron and transmission electron microscopy was fit to the $t^{1/3}$ power-law to determine coarsening rates and interface energy [10-13]. At the same time, direct, "bottom-up" approaches are also available to assess interface energies. In a model due to Bekker [14], the interface energy is estimated from the enthalpy difference between the two phases that create the interface, with a pre-factor dependent on the number of atoms and cross-bonds per unit area of the interface [14]. The widespread use [15-18] of the Bekker model stems from its simplicity, generality, and appeal to scientific intuition; however, it assumes simple, atomically planar interfaces, while not allowing for precipitate curvature, diffuse interfaces, compositional variations, or matrix non-crystallinity. While the Bekker model could be replaced by more recent methodologies for optimizing atomic structure and determining interfacial energy [19, 20], those methodologies may not be efficient when one side of the interface is amorphous. Furthermore, the use of density functional theory calculations improves the accuracy of the computed interface energy values for planar interfaces [21], but may not suitably address cases of curved interfaces, amorphous matrices, and compositional gradients.

In this article, we use atom-probe tomography (APT) and growth simulations to tackle the determination of interface energy of a crystalline-amorphous interface between precipitate and matrix, for the case of Cu clusters crystallized in an amorphous FINEMET-type alloy (composition $Fe_{72.89}Si_{16.21}B_{6.90}Nb_3Cu_1$). The fundamental importance of the Cu clusters is that they serve as nucleation sites for the larger $Fe_3Si$ nanoparticles [22, 23] that are responsible for the soft magnetic properties of annealed, nanocrystalline FINEMET [24-26]. At the interface of the Cu clusters with the Fe-Si-Nb-B-Cu matrix, the phases are dissimilar in both chemistry and structure: the matrix is amorphous and rich in Fe and Si, while the precipitates are crystalline Cu clusters that are neither planar nor regularly shaped [22, 23, 27]. As such, the Bekker model for interface energy [14] is not applicable, and accurate growth simulations of Cu precipitation in FINEMET have not been pursued so far. We carry out simulations of growth and coarsening via the Langer-Schwartz model [3, 5] (as implemented in the commercial software Thermocalc [18]) for a range of interface energy values, then compare the mean radius and size distribution obtained at each value of the interface energy with our experimental data. The interface energy governing the crystallization of Cu clusters in the Fe-Si-B-Nb-Cu system is that for which the mean radius and size distribution of simulated Cu clusters closely matches the APT data, $\gamma = 0.54$ J/m$^2$. The determined interface energy value differs by only about 15% from the Thermocalc default (Bekker model), but this difference triggers significant variations in the size distributions of Cu clusters. We analyze the driving force, nucleation rate, and volume fraction of Cu clusters, and make comparisons to other experimental works as well. These results highlight the sensitivity of the precipitation processes to the interface energy value, and the approach can be applied to other nanocrystals crystalized within amorphous matrices as well.

In our characterization procedure, amorphous FINEMET samples ($Fe_{72.89}Si_{16.21}B_{6.90}Nb_3Cu_1$, atomic %) were annealed at 540 °C for one hour, after which the specimens were prepared and analyzed using APT. The atom probe analysis was performed on a Cameca LEAP 4000X Si instrument using laser pulsing and a 90 mm flight path. The base temperature was 57.3 K. Laser pulsing was performed at 625 kHz using energies from 30 – 60 pJ at a detection rate of 6 to 30 ions per 1000 pulses. The standing bias during analysis was 1200 – 7500 V. Transmission electron microscopy was performed on the specimens before and after APT analysis, providing additional constraints for generating the reconstruction [28]. Figure 1(a) shows the specimen (tip) before APT. The size and number density of Cu clusters were determined employing the maximum separation method, using clusters with more than 35 Cu atoms separated by a maximum distance of 0.4 nm; these analysis parameters are comparable with a previous report on FINEMET with a somewhat different composition [29]. With the detection performed via the maximum separation method, the Cu clusters are depicted as orange domains in Figures 1(b) and 1(c) which show, respectively, a side view of the specimen and an axial view of a slice perpendicular to the tip axis. The average Cu cluster obtained from our APT reconstructions are approximately 2.5 nm in radius, a value consistent with previous reports [22, 27, 29, 30].

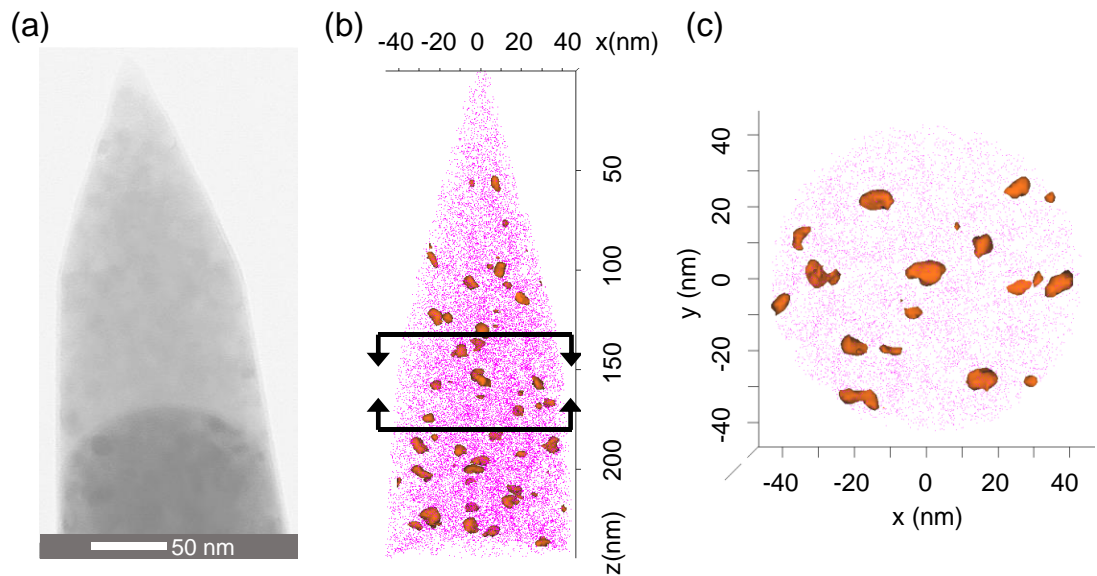

**Figure 1.** (a) FINEMET tip used for APT. (b) Three-dimensional APT reconstruction showing the Cu clusters colored orange (side view). An axial view of the tip domain delimited by the black bracket is shown in panel (c).

In addition to APT characterization, we also used a precipitation model developed in the TC-PRISMA module [6, 17] of Thermocalc [18] to simulate the growth of fcc-phase copper clusters [23] during isothermal annealing. The TC-Prisma module [6, 17] implements the Wagner-Kampmann numerical approach [3] for the Langer-Schwartz theory [5] to model the simultaneous nucleation, growth, and coarsening; this is a widely accepted technique that yields the mean radius of the nanocrystals and the particle size distribution as functions of time [4, 7, 8, 31]. We used the thermodynamic TCFE8 database [32] and the mobility MOBFE3 database [33], which have been shown to perform sufficiently well for more complex nanocrystalline phases [34-36]. The starting matrix is simulated as a non-crystalline material with the nominal composition $Fe_{72.89}Si_{16.21}B_{6.90}Nb_3Cu_1$ (atomic %) and density of 8.35 g/cm$^3$. Isothermal annealing was simulated at 540 °C. The interface energy $\gamma$ was varied in the range of 0.2 to 0.8 J/m$^2$, and for each value of $\gamma$ we analyzed the evolution of mean radius, size distribution, volume fraction, nucleation rate and number density obtained from the simulations.

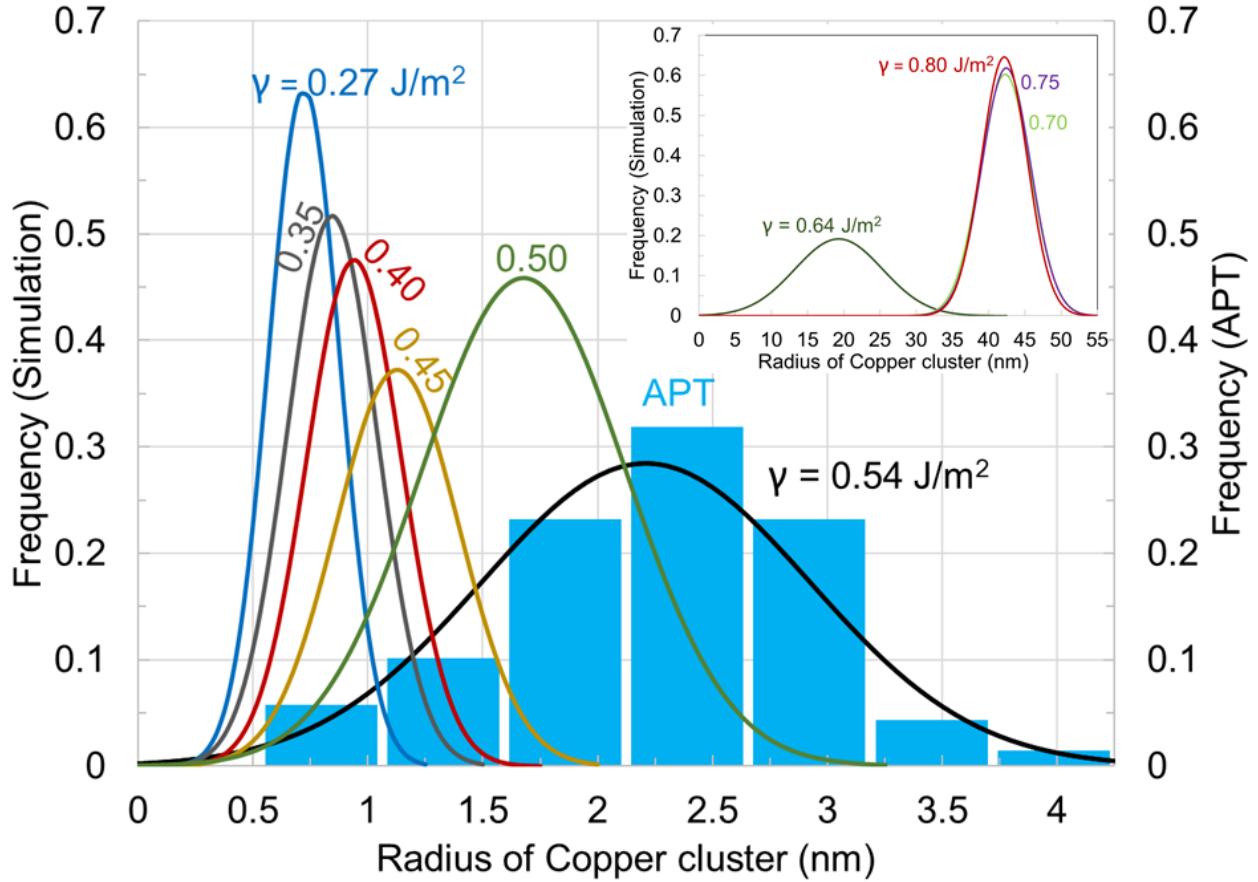

**Figure 2**. Cu cluster size distributions simulated for different values of the interface energy $\gamma$ (curves), compared with the distribution obtained from APT (blue histogram). The best match between APT results and the Cu cluster kinetics simulations is obtained a value of $\gamma$ = 0.54 J/m². The inset shows that for larger values of $\gamma$, the simulated Cu clusters are significantly fewer and grow excessively large.

Figure 2 shows a comparison of the size distribution of Cu clusters obtained from APT with the distributions from our coarsening simulations performed for different interface energies, at the same temperature (540 °C) and holding time (1 h). The APT data are from one specimen with 69 identified clusters. A second APT specimen run from the same sample shows a very similar distribution. As mentioned above, our cluster size distribution results are also consistent with the APT results previously found by others for similar alloys under similar processing [22, 27, 29, 30]. The best agreement of the simulations with our APT experiments is obtained for $\gamma$ = 0.54 J/m². We notice that the evolution during growth and coarsening is very sensitive to the interface energy value. The mean radius increases slowly for $\gamma$ values up to 0.58 J/m², after which the Cu clusters become very large (e.g., 20 nm radius at 0.64 J/m²), and eventually saturate at ~ 42 nm radius for $\gamma > 0.7$ J/m$^2$ (inset to Figure 2). Although it is tempting to rationalize analytically the dependence of mean radius on interface energy, the equations of growth and coarsening are complex (with many of the terms involved containing the interface energy) [3-8, 31] that such a rationalization would have a narrow scope. The default value for interface energy in TC-Prisma [17] is computed using the Bekker model [14] (0.64 J/m$^2$); while this is not too far off from the optimal value (0.54 J/m$^2$, Figure 2) obtained from comparisons with APT results, the entire evolution changes markedly, and the Cu clusters are one order of magnitude larger than those obtained in experiments. The results shown in Figure 2 therefore emphasize that the Bekker model [14] is in fact only a starting estimate that needs to be corrected when quantitative agreement with the experiments is sought.

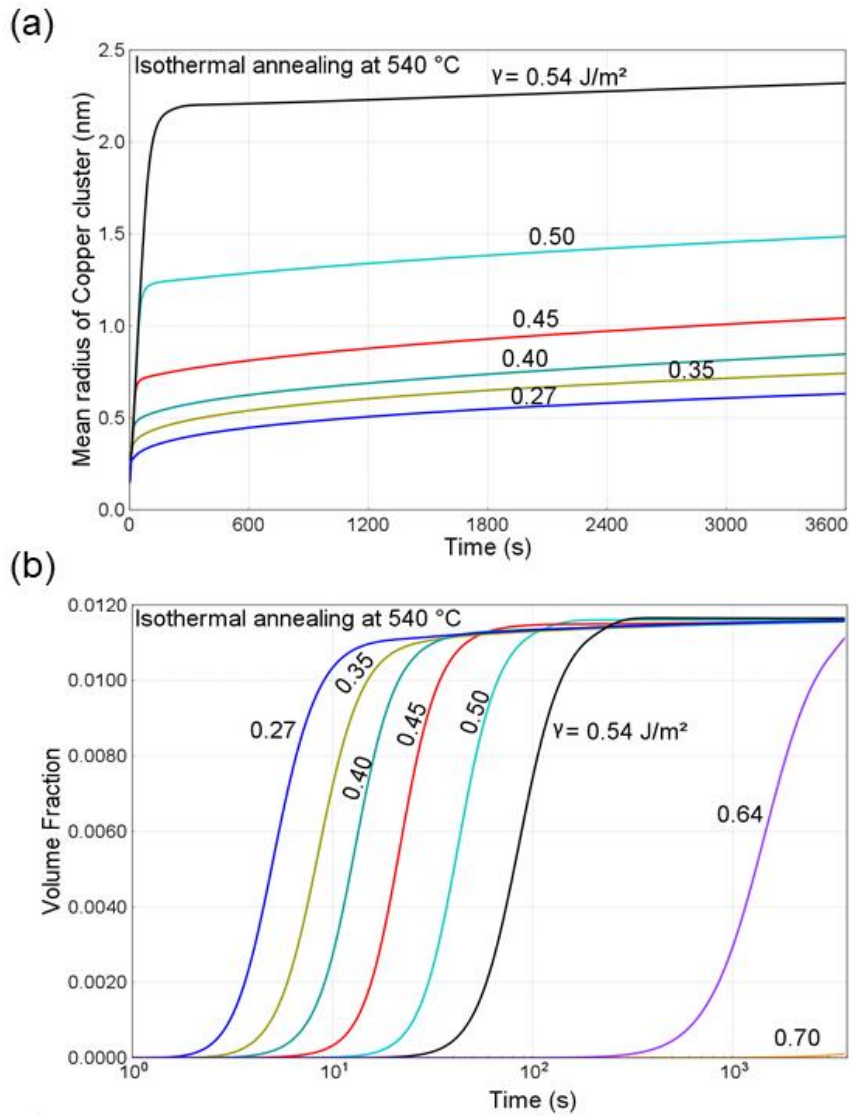

**Figure 3.** Mean radius (a) and volume fraction (b) of Cu clusters obtained from precipitation simulations with different interface energy values. The value determined from comparisons with APT results leads to saturation of the growth of Cu clusters after approximately 3 minutes.

Figure 3 shows the evolution of the mean radius and volume fraction as functions of time at several values of the interface energy. For $\gamma < 0.7$ J/m$^2$, the mean radius curves show an increasing trend as a function of $\gamma$ [Figure 3(a)], while the volume fraction saturates at ∼ 1.2% for all curves in Figure 3(b). Higher interface energies lead to excessive growth of the Cu clusters (inset to Figure 2), with negligible volume fraction [not shown in Figure 3(b)]. This excessive growth of a few clusters is consistent with our numerical observation that large interface energies decrease nucleation rates and driving forces by orders of magnitude (as compared with the results for $\gamma \leq 0.64$ J/m$^2$), hence leading to a few clusters large clusters instead of statistical size distributions (Figure 2, inset). At $\gamma = 0.54$ J/m$^2$, the Cu clusters nucleate rather rapidly in the simulations, in less than 2 minutes. This rapid crystallization is shown in Figure 3, and is consistent with the (computed) nucleation rate that drops a few orders of magnitude in the first 2 minutes, and vanishes altogether after about three minutes (Figure 4); the driving force has a similar behavior. The simulation results on nucleation rate and driving force (Figure 4) agree reasonably well with older experimental reports showing Cu clusters formed after 5 minutes [22]; the experiments were carried out at 450 ºC [22], and it is natural to expect that higher temperatures (such as those in our own experiments and simulation) lead to faster formation of the Cu clusters.

The number density of Cu clusters in our simulations stabilizes at $10^{24}$ m$^{-3}$ [Figure 4(b)], which is close to the density reported experimentally.[22, 37]

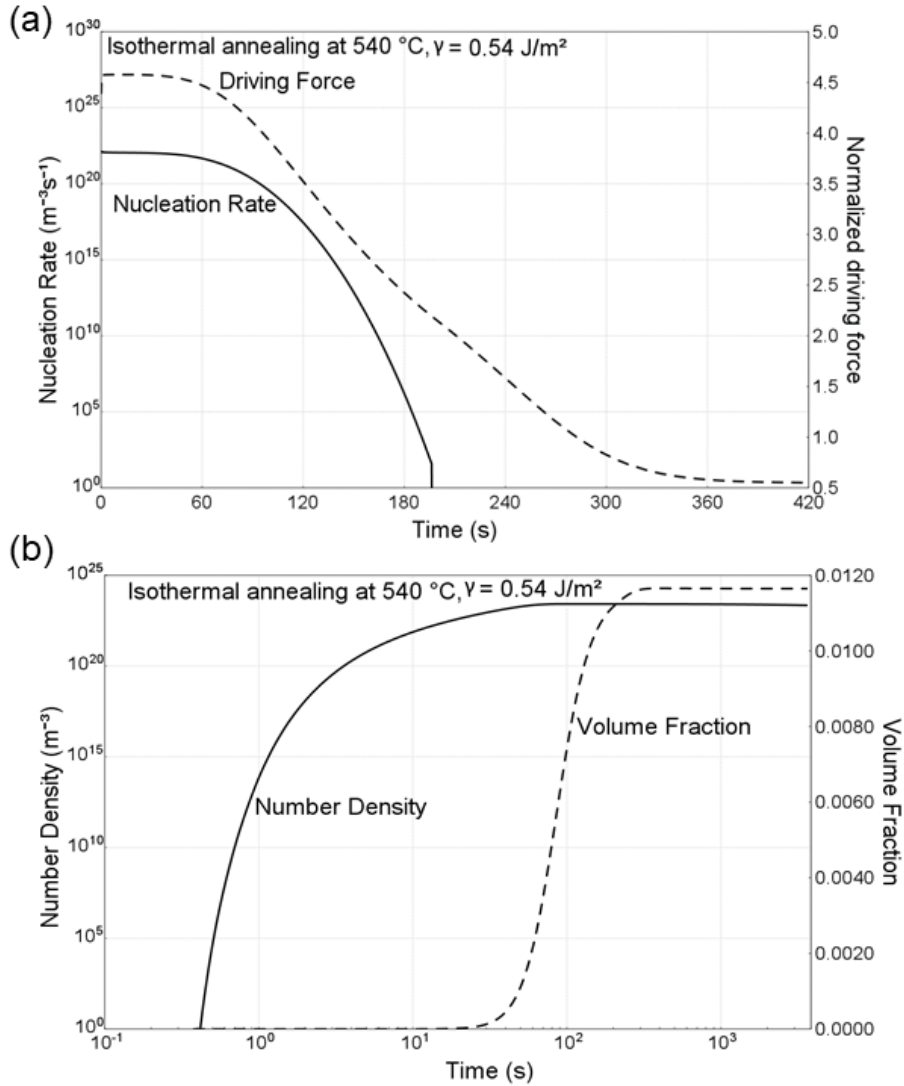

**Figure 4**. Computed driving force (a) and nucleation rate (b) of Cu clusters as a function of time during isothermal annealing at 540 °C. The calculations are carried out for $\gamma$ = 0.54 J/m$^2$, and show saturation behavior consistent with those of the mean radius and volume fraction in Figure 3.

There are several points resulting from the above APT-growth simulations determination of interfacial energy. From a fundamental point of view, it appears that the time evolution of the mean radius of Cu clusters [Figure 3(a)] for $\gamma$ > 0.40 J/m$^2$ does not follow the cube-root power law from the sharp interface [2, 13] or diffuse interface models [10-12]. Instead, two linear portions, one for growth and one for coarsening, seem to describe the evolution [Figure 3(a)]. It is not clear at present what the reason for this dependence is, but it may arise from the fact that the time interval of the simulation (which was chosen to match experiments) is too short to fully capture an asymptotic power-law coarsening behavior. Another point of note is that the value of the interface energy is an order or magnitude larger than that obtained for Ni-based alloys between the matrix and the γ' phase [12, 13]. This can be understood from the fact that the matrix and the precipitate in the case of Cu clusters in amorphous FINEMET are very different, hence many Cu atoms (per unit area) may be not fully bonded in the interface. This would be consistent with the fact that $Fe_3Si$ clusters nucleate in the proximity of Cu clusters, as the Cu precipitates would still be active, i.e. Cu atoms not fully passivated by the matrix. If desired, the density of Cu clusters can be controlled by the Cu content in the initial matrix, as well as by the density of the amorphous matrix itself. Since the Cu clusters serve for the nucleation of magnetic D03 phase, Ohnuma et al. [38] have investigated the optimization of Cu concentration in FINEMET-type alloys so that the magnetic permeability of the alloy is maximized after the secondary crystallization. In this respect, the Cu clusters should have sufficient number density, while also being large enough to serve as nucleation centers for the D03 phase [38]. Thermocalc simulations with the correct interface energy can be carried out for target number density and mean size or Cu clusters, and thus could subsequently aid for magnetic property optimization.

In conclusion, we have used APT and parametric simulations of growth in Thermocalc to determine an interface energy of 0.54/ $m^2$ for Cu crystalline clusters in amorphous FINEMET. In the process, we have shown that the growth of Cu clusters is highly sensitive to the interface energy value, and that the currently used Bekker model [14] for interface energy does not correctly capture the growth of Cu clusters in FINEMET. The interface energy value determined from this combined APT-simulation approach can be used in future simulation efforts to guide the optimization of soft magnetic alloys. For example, such simulations should aim to elucidate the effect of initial Cu concentration in Fe-Si-B-Nb-Cu and/or the effect of ageing time during the Cu (primary) precipitation on the secondary precipitation of $Fe_3Si$ nanocrystals that are responsible for the soft-magnetic behavior; the latter may involve exploring initial growth behavior, i.e., outside the asymptotic power law regime. Ultimately, exploring the possibility of forming primary nucleation sites from components other than Cu and/or alloys other than FINEMET, will require this experimental-simulation approach to first determine the interface energy value, and subsequently use it in simulations to understand how composition, temperature and aging time control the size and density of nucleated clusters.

**Data Availability:** Figures and data created during this work have been deposited at Citrine Informatics, as the set labelled DATA: Interface Energy of Cu clusters in FINEMET (2018); files are available for public access at https://citrination.com/datasets/157196/show_files.

## ACKNOWLEDGEMENT

The authors gratefully acknowledge support from the National Science Foundation through Granthttps://citrination.com/datasets/157196/show_files No. DMREF-1629026. We thank Prof. Matthew Willard (Case Western Reserve University), as well as Dr. Alex Leary and Dr. Ronald Noebe (NASA Glenn Research Center) for useful discussions and initial critiques of this work.

## REFERENCES

[1]  C. Wagner, Z. Elektrochemie 65 (1961) 581–594.
[2]  I. M. Lifshitz and V. V. Slyozov, J. Phys. Chem. Solids 19 (1961) 35–50.
[3]  R. Wagner and R. Kampmann, in Materials Science and Technology - A Comprehensive Treatment, Vol. 5, VCH, Weinheim  (1991) p. 21.
[4]  M. Perez, M. Dumont, and D. Acevedo-Reyes, Acta Mater. 56 (2008) 2119–2132.
[5]  J. S. Langer and A. J. Schwartz, Phys. Rev. A 21 (1980) 948–958.
[6]  Q. Chen, K. S. Wu, G. Sterner, and P. Mason, J. Mater. Eng. Perform. 23 (2014) 4193–4196.
[7]  O. R. Myhr and O. Grong, Acta Mater. 48 (2000) 1605–1615.
[8]  J. D. Robson, Acta Mater. 52 (2004) 4669–4676.
[9]  H. A. Calderon, P. W. Voorhees, J. L. Murray, and G. Kostorz, Acta Metall. Mater. 42 (1994) 991–1000.
[10]  A. J. Ardell, Acta Mater. 58 (2010) 4325–4331.
[11]  A. J. Ardell, Acta Mater. 61 (2013) 7828–7840.
[12]  A. J. Ardell, J. Mater. Sci. 46 (2011) 4832–4849.
[13]  S. Meher, M. C. Carroll, T. M. Pollock, and L. J. Carroll, Mater. Des. 140 (2018) 249–256.
[14]  R. Bekker, Ann. Phys. 424 (1938) 128–140.
[15]  T. Nishizawa, I. Ohnuma, and K. Ishida, J. Phase Equilib. 22 (2001) 269–275.
[16]  X. Li, N. Saunders, and A. P. Miodownik, Metall. Mater. Trans. A 33A (2001) 3367–3373.
[17]  http://www.thermocalc.com/products-services/software/precipitation-module-(tc-prisma); http://www.thermocalc.com/media/36120/TC-Prisma-User-Guide_2015a.pdf.
[18]  http://www.thermocalc.com/.
[19]  C. V. Ciobanu, V. B. Shenoy, C. Z. Wang, and K. M. Ho, Surf. Sci. 544 (2003) L715–L721.
[20]  C. V. Ciobanu and C. Predescu, Phys. Rev. B 70 (2004) Art. no. 085321.

[21] M. Y. Yu, W. W. Qu, S. M. Xu, L. Wang, B. G. Liu, L. B. Zhang, and J. H. Peng, Comp. Mat. Sci. 153 (2018) 217–227.
[22] K. Hono, D. H. Ping, M. Ohnuma, and H. Onodera, Acta Mat. 47 (1999) 997–1006.
[23] K. Hono, Prog. Mater. Sci. 47 (2002) 621–729.
[24] M. E. McHenry, M. A. Willard, and D. E. Laughlin, Prog. Mater. Sci. 44 (1999) 291–433.
[25] G. Herzer, IEEE Trans. Magn. 26 (1990) 1397–1402.
[26] G. Herzer, Scripta Metall. Mater. 33 (1995) 1741–1756.
[27] K. Hono and D. H. Ping, Mater. Charact. 44 (2000) 203–217.
[28] B. P. Gorman, A. Puthucode, D. R. Diercks, and M. J. Kaufman, Mater. Sci. Technol. 24 (2008) 682–688.
[29] K. G. Pradeep, G. Herzer, P. Choi, and D. Raabe, Acta Mater. 68 (2014) 295–309.
[30] K. Hono, J. L. Li, Y. Ueki, A. Inoue, and T. Sakurai, Appl. Surf. Sci. 67 (1993) 398–406.
[31] J. D. Robson and P. B. Prangnell, Acta Mater. 49 (2001) 599–613.
[32] http://www.thermocalc.com/media/10306/dbd_tcfe8_extendedinfo.pdf.
[33] http://www.thermocalc.com/media/32320/marketing-database-overview-oct-2016.pdf.
[34] R. Jha, N. Chakraborti, D. R. Diercks, A. P. Stebner, and C. V. Ciobanu, Comp. Mater. Sci. 150 (2018) 202–211.
[35] H. W. Yen, M. H. Chiang, Y. C. Lin, D. Chen, C. Y. Huang, and H. C. Lin, Metals 7 (2017).
[36] A. K. Behera and G. B. Olson, Scripta Mater. 147 (2018) 6.
[37] G. M. Li, D. R. Li, X. J. Ni, Z. Li, and Z. C. Lu, Rare Metal Mat. 42 (2013) 1352–1355.
[38] M. Ohnuma, K. Hono, S. Linderoth, J. S. Pedersen, Y. Yoshizawa, and H. Onodera, Acta Mat. 48 (2000) 4783–4790.